\input{aipcheck}

\documentclass[
    ,final            
  ]
  {aipproc}

\layoutstyle{6x9}

\begin{document}

\title{Nonperturbative Physics in a Magnetic Field}


\author{Vivian de la Incera}{address={Department of Physics\\
The University of Texas at El Paso\\
El Paso,
TX 79968,
USA}}

\begin{abstract}
Non-Perturbative Quantum Field Theory has played an important role in the study of phenomena where a fermion condensate can appear under certain physical conditions. The familiar phenomenon of electric superconductivity, the color superconductivity of very dense quark matter, and the chiral symmetry breaking of low energy effective chiral theories are all examples of that sort. Often one is interested in the behavior of these systems in the presence of an external magnetic field. In this talk I will outline the effects of an external magnetic field on theories with either fermion-fermion or fermion-antifermion condensates.
\end{abstract}

\maketitle


\section{Dynamical Symmetry Breaking}

 There are many important physical phenomena that cannot be described with the help of a perturbative expansion. Notorious examples are superconductivity, chiral symmetry breaking, and the formation of mesons out of quarks and antiquarks. They are all realizations of the phenomenon of dynamical symmetry breaking (DSB).  The main purpose of this talk is to review some of the key effects of an external magnetic field in systems with DSB. For the sake of understanding however, it is better to start summarizing the basic ingredients of the phenomenon of DSB in the absence of external magnetic fields.

 Dynamical symmetry breaking is a special case of \emph{Spontaneous Symmetry Breaking}. Spontaneous symmetry breaking (SSB) occurs when the vacuum or ground state of a system does not posses the symmetries of the Hamiltonian.  There are many non-relativistic and relativistic systems with SSB. A well-known non-relativistic example with SSB is the Heisenberg ferromagnet. It consists of an infinite crystalline array of spin-$\frac{1}{2}$ magnetic dipoles characterized by a rotationally invariant Hamiltonian. However, the spin-spin interactions between nearest neighboring dipoles tend to align them in some arbitrary direction. There is no preference for any particular alignment direction, so all the possible alignments have the same energy and thus the ground state is degenerate. Any given alignment spontaneously breaks the rotational symmetry of the Hamiltonian. A similar situation occurs in relativistic field theories. A classical example is a theory of n elementary real scalar fields $\phi$ with Lagrangian density $L=\frac{1}{2}\partial_{\mu}\phi\partial^{\mu}\phi-V(\phi)$. We can assume that the potential is such that the Lagrangian is invariant under some symmetry transformation. If the potential can be written in the form $V(\phi)=\frac{\lambda}{4}(\phi^{2}-a)^{2}$, the theory is invariant under the discrete transformation $\phi\rightarrow-\phi$, but the potential has two equivalent minima (or vacuum expectation values) at $<\phi>= a$ and $-a$. There is no preference for any one of them, but once one is chosen, the symmetry is spontaneously broken. Notice that a common element of these two examples is the existence of a degenerate vacuum or ground state. This is a key feature for a system with SSB.

 We are interested here in relativistic field theories, so from now on we concentrate on them. One should distinguish between two classes of SSB, depending on what kind of symmetry is broken: discrete or continuous. The relativistic example we just mentioned exhibited SSB of a discrete symmetry. However, the SSB of a continuous symmetry is particularly interesting because it almost always implies the existence of massless spinless particles, known as Goldstone bosons. This is the famous Goldstone theorem \cite{Goldstonetheo}. We say \emph{almost always} because this is only true if the theory obeys certain axioms like Lorentz invariance, locality, etc. (see the book of Coleman \cite{Colemanlectures} for an excellent discussion of SSB and a very intuitive explanation of the Goldstone theorem). Most of the field theories obey these axioms. Therefore, the appearance of massless spinless particles is a hallmark of spontaneously broken continuous symmetries.  In the context of dynamical symmetry breaking, the Goldstone (or pseudoGoldstone) bosons, being the light particles, are the relevant degrees of freedom in the low-energy dynamics of effective theories with SSB.

 The case of gauge field theories (theories with massless vector fields that are invariant under some local continuous symmetry transformation) is exceptional in the sense that there is no gauge on which all the axioms of the Goldstone theorem can be satisfied. Hence, in the case of a gauge theory no Goldstone bosons are actually produced when SSB takes place. Instead, some of the originally massless vector mesons become massive thanks to the SSB. One can casually say that these vector fields "eat up" the Goldstone bosons in order to acquire a mass and then carry three instead of the two degrees of freedom they would have had as massless vector fields. The precise mechanism by which this can be realized is the Higgs mechanism. The Higgs mechanism, which was actually developed by several different authors \cite{Higgsmechanism}, is based on gauge theories that contain elementary scalar fields whose nonzero vacuum expectation values spontaneously break the symmetry of the action giving mass to vector mesons and fermions that interact with them. The implementation of the Higgs mechanism involve shifting the scalar field by his vacuum expectation value (VEV),  $\phi=\phi'+ <\phi>$, and rewriting the Lagrangian in terms of the new (primed) fields which have zero VEV. This process gives rise to new mass terms and new interactions among the fields. The details can be found in any book on quantum field theory, so we are not going to extend on this point any longer. The relevance of the Higgs mechanism is that it provides a way to give mass to vector mesons in non-Abelian gauge theories without sacrificing the renormalizability of the theory.

Giving mass to particles through the conventional Higgs mechanism requires the presence of scalar fields whose expectations values can be nonzero.  However, to date and despite innumerable attempts, no elementary scalar (Higgs) field has been detected in the experiment, although the LHC is aimed to provide a definite resolution to this quest in the near future. This limitation led to the pioneers works of Nambu and Jona-Lasinio \cite{NambuJL1961}, who showed that a theory without fundamental scalar fields may possess symmetry-violating solutions thanks to the presence of attractive four-fermion interactions. This interaction led to a fermion-antifermion condensate and to a bound-state Goldstone boson. Their model was the first example of a system undergoing \emph{dynamical symmetry breaking}, which is nothing but a special case of spontaneous symmetry breaking in which the order parameter is given by the vacuum average of a composite, instead of single, operator. Nambu and Jona-Lasinio work inspired other authors to explore this possibility further to include dynamical breaking of gauge symmetries \cite{dsbgaugetheories}.

Later on, a formalism to systematically analyze DSB with the help of functional methods was developed in the seminal paper of Cornwall, Jackiw and Tomboulis \cite{CJT1974}. Their method is based on the so-called effective action of composite operators $\Gamma(\phi_{c},G)$. This functional is a generalization of the usual effective action $\Gamma(\phi_{c})$ in the sense that it is a functional of both the expectations values of elementary fields $\phi_{c}(x)=<\phi(x)>$ and of the propagator $G(x,y)=<0\mid\phi(x)\phi(y)\mid0>$. $\Gamma(\phi_{c},G)$ is the generating functional of two-particle irreducible n-point functions \cite{CJT1974}. The effective action of composite operators formalism is a nonperturbative scheme, as it includes the resummation of an infinite number of diagrams. The topic of nonperturbative physics and theories with DSB is vast and we do not have space in a conference proceeding paper to provide a comprehensive list of references, so we refer the interested reader to the book of Miransky \cite{Miranskybook} and  references therein.

What happens to DSB in the presence of an external magnetic field? It depends on which realization of DSB one is considering. A strong enough external magnetic field may influence DSB in quite different ways.  It may just destroy DSB, as it happens in the case of type I electrical superconductivity, it may actually trigger DSB, as in the phenomenon of Magnetic Catalysis of Chiral Symmetry Breaking \cite{MC}-\cite{F&I:MC98}; it may change the symmetry of the ground state, as it does in the magnetic phases of color superconductivity \cite{FIM:05:06}-\cite{F&I:JPA}, or it may enhance the DSB, as in the case of chiral symmetry breaking in QCD \cite{Shupanov&Smilga97}. The rest of this paper will be dedicated to discussing the realization of DSB in the presence of an external magnetic field.

\section{Magnetic Catalysis of Chiral Symmetry Breaking}

\subsection{Early Approach}

Let us consider a theory of interactive massless fermions at zero density. In the presence of an external magnetic field and whenever there are attractive channels in the fermion interaction, the system will exhibit the phenomenon of magnetic catalysis of chiral symmetry breaking ($M\chi CSB$) \cite{MC}-\cite{MC-QED}, that is, the dynamical generation of a chiral condensate (and consequently of a fermion mass) in the presence of a magnetic field. A most significant feature of the $M\chi CSB$ is that it requires no critical value of the fermion's coupling for the condensate to be generated. That is, the symmetry breaking can take place at the weakest attractive interaction. Physically, what happens is that the magnetic field forces the low energy fermions to reside basically in their lowest Landau level (LLL). This, in turn, yields a dimensional reduction of the infrared fermion dynamics. The dimensional reduction is reflected in an effective strengthening of the fermion interactions leading to dynamical symmetry breaking through the generation of a fermion-antifermion condensate, and consequently, of a dynamical fermion mass \cite{MC}-\cite{F&I:MC98}.

The magnetically catalyzed dynamical mass can be found by solving the gap equation with the help of a non-perturbative approach like for instance the one described in \cite{CJT1974}. Using this method one can derive the Schwinger-Dyson equations for the fermion mass operator by minimizing the effective action of composite operators of the given theory with respect to the fermion propagator and then solving it within some nonperturbative approximation, for example the quenched ladder approximation.

\subsection{Magnetic Catalysis Revisited}

Until recently, all the studies of $MC$$\chi$$SB$ focused on the generation of a dynamical mass \cite{MC}-\cite{MC-Gen}, but
ignored the possibility of a dynamically generated anomalous magnetic moment. However, as shown in \cite{newMC}, once the chiral symmetry is broken, both a dynamical mass and a dynamical anomalous magnetic moment (AMM) can be generated. Physically it is easy to grasp the origin of this new dynamical quantity. The chiral condensate carries non-zero magnetic moment, since the particles forming the condensate have opposite spins and opposite charges. Therefore, chiral condensation will inexorably provide the quasiparticles with both a dynamical mass and a dynamical magnetic moment. Symmetry arguments can also help to understand this phenomenon. A magnetic moment term does not break any additional symmetry that has not already been broken by the dynamically generated mass. Hence, once $M\chi CSB$ occurs, there is no reason why only one of these parameters should be different from zero.

In Ref. \cite{newMC} the realization of $M\chi CSB$ in massless QED was reconsidered. To incorporate the possibility of a dynamically generated AMM, the structure of the fermion self-energy $\Sigma(x,x')$ was proposed to be of the form

\begin{equation}
\Sigma (x,x')=(Z_{\|}\Pi_\mu^{\|}
\gamma^\mu_{\|}+Z_{\bot}\Pi_\mu^{\bot} \gamma^\mu_{\bot} +M+\frac{T}{2}
\widehat{F}^{\mu \nu}\sigma_{\mu \nu})\delta^4(x-x')\label{Mass
Operator}
\end{equation}
Here $\Pi_\mu^{\|}$ and $\Pi_\mu^{\bot}$ are the parallel and transverse covariant derivatives; the wave function's renormalization coefficients $Z_{\|}$ and $Z_{\bot}$, the mass $M$ and  the anomalous magnetic moment $T$ are all physical parameters that have to be determined from the SD equations of the theory.

In (\ref{Mass
Operator}) $\widehat{F}^{\mu \nu}=F^{\mu
\nu}/H$ denotes the normalized electromagnetic strength tensor, with
$H$ the field strength. The external magnetic field breaks the rotational
symmetry of the theory, hence separating between longitudinal $p_{\|}\cdot \gamma^{\|}=p^{\nu }\widehat{F}_{\nu \rho}^{\ast} \widehat{F}^{\ast \mu \rho}\gamma _{\mu }$ (for $\mu, \nu=0,3$),
and transverse $p_{\bot}\cdot \gamma^{\bot}=p_{\mu}\widehat{F}^{\mu\rho}\widehat{F}_{\rho\nu}\gamma^{\nu}$ (for $\mu, \nu=1,2$) modes. $\widehat{F}_{\mu\nu}^{ \ast} =\frac{1}{2H}\varepsilon _{\mu \nu \rho \lambda}F^{\rho \lambda }$ is
the dual of the normalized electromagnetic strength tensor
$\widehat{F}_{\mu\nu }$.

The transformation to  momentum space of (\ref{Mass Operator}) can
be done with the help of Ritus' eigenfunctions. This method was originally developed
for fermions in \cite{Ritus:1978cj} and later extended to vector
fields in \cite{efi-ext}. In Ritus' approach, the transformation to
momentum space is carried out using the eigenfunctions
$E_{p}^{l}(x)$ of the asymptotic states of the charged fermions in a
uniform magnetic field
\begin{equation}\label{Ep}
 E_{p}^{l}(x)=E_{p}^{+}(x)\Delta(+)+E_{p}^{-}(x)\Delta(-)
\end{equation}
where
\begin{equation}
\Delta(\pm)=\frac{I\pm i\gamma^{1}\gamma^{2}}{2},
\label{Spin-projectors}
\end{equation}
are the spin up ($+$) and down ($-$) projectors, and
$E_{p}^{+/-}(x)$ are the corresponding eigenfunctions
\begin{eqnarray}\label{E-x}
E_{p}^{+}(x)=N_{l}e^{-i(p_{0}x^{0}+p_{2}x^{2}+p_{3}x^{3})}D_{l}(\rho),\qquad
\nonumber
\\
E_{p}^{-}(x)=N_{l-1}e^{-i(p_{0}x^{0}+p_{2}x^{2}+p_{3}x^{3})}D_{l-1}(\rho)
\end{eqnarray}
with normalization constant $N_{l}=(4\pi eH)^{1/4}/\sqrt{l!}$. The functions
$D_{l}(\rho)$ denote the parabolic cylinder functions of argument
$\rho=\sqrt{2eH}(x_{1}-p_{2}/eH)$ and Landau
level index $l=0,1,2,...$.

The $E_p^l$ functions satisfy the orthogonality condition
\cite{Wang}
\begin{equation}
\int d^{4}x \overline{E}_{p}^{l}(x)E_{p'}^{l'}(x)=(2\pi)^4
\widehat{\delta}^{(4)}(p-p')\Pi(l) \ , \label{orthogonality}
\end{equation}
with $\overline{E}_{p}^{l}\equiv \gamma^{0}
(E_{p}^{l})^{\dag}\gamma^{0}$,
\begin{equation}
\widehat{\delta}^{(4)}(p-p')=\delta^{ll'} \delta(p_{0}-p'_{0})
\delta(p_{2}-p'_{2}) \delta(p_{3}-p'_{3}), \label{gen-delta}
\end{equation}
and
\begin{equation}
\Pi(l)=\Delta(+)\delta^{l0}+I(1-\delta^{l0}). \label{degeneracy}
\end{equation}

In momentum space the self-energy becomes
\begin{eqnarray}\label{P-Self-Energy}
\Sigma(p,p')= \int d^4xd^4y
\overline{E}_{p}^{l}(x)\Sigma(x,y)E_{p}^{l}(y)=(2\pi)^4\widehat{\delta}^{(4)}(p-p')\Pi(l)\widetilde{\Sigma}^l
(\overline{p})
\end{eqnarray}
with
\begin{equation}\label{SE-LLL}
\widetilde{\Sigma}^{l}(\overline{p})
=Z_{\|}^{l}\overline{p}_{\|}^\mu\gamma_{\mu}^{\|}+Z_{\bot}^{l}\overline{p}_{\bot}^\mu\gamma_{\mu}^{\bot}+M^{l}I+iT^{l}\gamma^{1}\gamma^{2},
\end{equation}
$\overline{p}_{\|}^\mu =(p_{0},0,0,p_{3})$ and $\overline{p}_{\bot}^\mu=(0,0,\sqrt{2eHl},0)$.

Using this self-energy, one can show \cite{newMC} that the corresponding Schwinger-Dyson equation in the quenched ladder approximation can be written as
\begin{eqnarray}\label{SD-Eq}
\widetilde{\Sigma}^{l}(\overline{p})\Pi(l)=
-ie^2(2eH)\Pi(l)\int\frac{d^4\widehat{q}}{(2\pi)^4}
\frac{e^{-\widehat{q}^2_\bot}}{\widehat{q}^2}[\gamma_{\mu}^{\|}\widetilde{G}^{l}(\overline{p-q})\gamma_{\mu}^{\|}
\nonumber
\\
+\Delta(+)\gamma_{\mu}^{\bot}\widetilde{G}^{l+1}(\overline{p-q})\gamma_{\mu}^{\bot}\Delta(+)
+
\Delta(-)\gamma_{\mu}^{\bot}\widetilde{G}^{l-1}(\overline{p-q})\gamma_{\mu}^{\bot}\Delta(-)]
\end{eqnarray}

where
\begin{eqnarray}\label{Pgorrito}
\widetilde{G}^{l}(\overline{p})
=\frac{N^l(T,V_{\|})}{D^l(T)}\Delta(+)\Lambda^{+}_{\|}
+\frac{N^l(T,-V_{\|})}{D^l(-T)}\Delta(+)\Lambda^{-}_{\|}\qquad\qquad\qquad
\nonumber
\\
+\frac{N^l(-T,V_{\|})}{D^l(-T)}\Delta(-)\Lambda^{+}_{\|}
+\frac{N^l(-T,-V_{\|})}{D^l(T)}\Delta(-)\Lambda^{-}_{\|}\qquad\qquad\qquad\nonumber
\\
-iV_{\bot}^l(\Lambda^{+}_{\bot}-\Lambda^{-}_{\bot})
[\frac{\Delta(+)\Lambda^{+}_{\|}+\Delta(-)\Lambda^{-}_{\|}}{D^l(T)}
+
\frac{\Delta(+)\Lambda^{-}_{\|}+\Delta(-)\Lambda^{+}_{\|}}{D^l(-T)}]\qquad\qquad
\end{eqnarray}
with notation
\begin{eqnarray}\label{coefficients}
N^l(T,V_{\|})&=&T^l-M^l-V_{\|}^l \qquad \qquad \nonumber
\\
D^l(T)&=&(M^l)^2-(V_{\|}^l-T^l)^2+(V_{\bot}^l)^2 \qquad \nonumber
\\
V_{\|}^l&=&(1-Z_{\|}^{l})|\overline{p}_{\|}|\qquad \nonumber
\\
V_{\bot}^l&=&(1-Z_{\bot}^{l})|\overline{p}_{\bot}|=(1-Z_{\bot}^{l})\sqrt{2eHl}.\qquad
\end{eqnarray}

As the fermions in the LLL have only one spin orientation it is not
possible to find $M^{0}$ and $T^{0}$ independently. That means that in the $LLL$ the SD equation (\ref{SD-Eq}) can only determine
the induced LLL rest-energy $E^{0}=M^{0}+T^{0}$. The result for this combination is \cite{newMC}
\begin{equation}
\label{Mass-Eq-Solution} M^{0}+T^{0}\simeq \sqrt{2eH}
e^{-\sqrt{\frac{\pi}{\alpha}}}
\end{equation}

For fermions in the first LL one has \cite{newMC}
\begin{equation}\label{M}
M^{1}=-T^{1}=\frac{1}{2} E^{0}=\sqrt{eH/2}
e^{-\sqrt{\frac{\pi}{\alpha}}},\qquad
\end{equation}
corroborating the relevance of the LLL dynamics (both
$M^{1}$ and $T^{1}$ are determined by $E^{0}$) in the generation of
the dynamical mass and magnetic moment of the fermions in the first
LL. The dynamically induced $T^{1}$ produces the energy splitting
\begin{equation}\label{energy-splitting}
\Delta E=|2T^{1}|=2 \sqrt{eH/2} e^{-\sqrt{\pi /\alpha}},
\end{equation}
which can be rewritten in the usual form of the Zeeman
energy splitting for the two spin projections:
$\Delta E=\widetilde{g}\widetilde{\mu}_B H$
with $\widetilde{g}$ and $\widetilde{\mu}_B$ representing the
non-perturbative Lande $g$-factor and Bohr magneton respectively given by
$\widetilde{g}=2e^{-2\sqrt{\pi /\alpha}},\quad \widetilde{\mu}_B
=\frac{e}{2M^1}$.
This splitting is nothing but a dynamically induced Zeeman effect characterized by a Lande g-factor $\widetilde{g}=2e^{-2\sqrt{\pi /\alpha}}$ which depends nonperturbatively on the coupling constant, and a Bohr magneton $\widetilde{\mu}_B
=\frac{e}{2M^1}$ that is given in terms of the dynamically induced electron mass, in analogy to the way in which the bare Bohr magneton depends on the bare mass.

\section{Color Superconductivity in a Magnetic Field}
Let us consider now the effects of an external magnetic field on theories of interacting quarks at high baryonic density. At densities of the order of $10$ times the nuclear density the
quarks in baryonic matter will be so weakly interacting that they can exist out of confinement. What is
particularly interesting about cold and dense quark matter is that the fundamental
QCD interaction is attractive in the color antitriplet channel. Once the quarks are
deconfined and fill out the available quantum states up to the Fermi surface, this
attractive interaction triggers the formation of diquark pairs at the Fermi surface, leading to the phenomenon of color superconductivity (CS) (see the reviews on this topic \cite{reviews} and reference therein).
In nature the combination of the high densities and relatively low temperatures
required for color superconductivity can be found in the interior of neutron stars. On the other hand, magnetic fields as large as $10^{12}-10^{13} G$ exist in the surface of regular neutron stars. In the case of magnetars they are in the range of $10^{14}-10^{15}G$, and perhaps as high as $10^{16}G$. The star's interior field can be several orders of magnitude larger. To produce reliably predictions
of astrophysical signatures of color superconductivity, a better understanding of the role of the star's magnetic field in the CS phase is essential.

In recent years, several works \cite{FIM:05:06}-\cite{F&I:JPA} have been dedicated to elucidate the influence
of a magnetic field in the ground state of CS matter. These investigations have
revealed a richness of phases \cite{MPhases} with different symmetries and low energy properties.
In order to grasp how a magnetic field can affect the color superconducting pairing,
it is important to recall that in spin-zero color superconductivity, although the color
condensate has non-zero electric charge, a linear combination of the photon and one
of the gluons remains massless, so the condensate is neutral with respect
to the Abelian charge associated with the symmetry group of this long-range gauge
field. This combination then behaves as the "in-medium" (also called "rotated")
electromagnetic field in the color superconductor. Since this combination acquires
no mass, there is no Meissner effect for the corresponding "rotated" magnetic field
and consequently, a spin-zero color superconductor may be penetrated by a
magnetic field B. Moreover, it is worth to notice that even though all the superconducting
pairs are neutral with respect to this long-range field, a subset of them is formed by
quarks of opposite rotated charges. The interaction of the charged quarks with the
magnetic field gives rise to a difference between the gaps getting contribution from
pairs formed by oppositely charged quarks and those getting contribution only from
pairs of neutral quarks. One consequence of such a difference is the change of the
gap parameter symmetry \cite{FIM:05:06}. If the field is strong enough, it actually strengthens the
pairing of quarks of oppositely rotated charge. One can intuitively understand
this considering that the quarks with opposite rotated charges and opposite spins, have
parallel (rather than antiparallel) magnetic moments, so the field tends to keep the
alignment of these magnetic moments, hence helping to stabilize the pairing of these
quarks.

Besides changing the symmetry of the gap and consequently the low-energy physics,
a magnetic field can also lead to other interesting behaviors. It can produce oscillations
in the gaps and the magnetization \cite{MOscillations}, the Hass-Van Alphen effect. Moreover,
when the field strength is of the order of the Meissner mass of the rotated charged
gluons, these modes become tachyonic \cite{F&I:PRL06}. The solution to this instability is the
formation of a vortex state of gluons which in turn boosts the magnetic field, creating
a peculiar paramagnetic state. This magnetic-field induced gluon vortex state is
known as the Paramagnetic CFL (PCFL) phase.

A magnetic field, or more precisely, the spontaneous generation of one, could be related to the most urgent question in the field of color superconductivity, i.e. the determination of the stable ground state that realizes at the intermediate densities present at the cores of neutron stars. At intermediate densities, when the Fermi surface imbalance between pairing quarks becomes too large, some of the modes become gapless \cite{gaplessCS}. The imbalance itself comes from imposing beta equilibrium and neutrality or when the density is such that the strange quark mass cannot be ignored any longer. It has been shown \cite{ChromoInstab} that a consequence of the gapless modes is the occurrence of chromomagnetic instabilities, a sign that one is not working in the correct ground state. In \cite{B-induction} a solution  to this chromomagnetic instability was considered. It involved the generation of an inhomogeneous gluon condensate and an induced magnetic field, so this kind of solution could also serve as a new mechanism to generate magnetic fields in the core of a neutron star \cite{F&Imagnetars}.

\section{Outlook}

In this paper we discussed the realization of DSB in a magnetic field at zero density: the phenomenon of magnetic catalysis of chiral symmetry breaking, and at finite density: the phenomenon of color superconductivity in a magnetic field. Even though the fermion condensates that characterize each of these phenomena are very different; one is a chiral condensate and it occurs at the Dirac sea, while the other is a fermion-fermion condensate that realizes at the Fermi surface, they have several common elements; they  both can be studied with the same nonperturbative methods and they both are mechanisms to give mass to fermions.

Now, for the $M\chi CSB$ we saw that a new dynamical parameter, the AMM, is generated along with the dynamical fermion mass once the chiral condensate is formed. The dynamical generation of an AMM may be relevant for the physics of condensed matter systems like graphene that can be described by two-dimensional Dirac-like Hamiltonians, so more work needs to be done to explore this phenomenon within different massless fermion models. On the other hand, there are many pending questions about the implications of a strong magnetic field in color superconductivity. Could the analogous of the dynamical AMM be also generated in the color superconductor in a magnetic field? After all, some of the Cooper pairs in the color superconductor are formed by quarks with opposite rotated charges and these pairs will carry a nonzero magnetic moment. What would be the implications of this new dynamical parameter for the superconductor? Could a dynamical AMM modify the spectrum of the quasiparticles and affect the onset of gapless modes in spin-zero color superconductivity? Would any of these things lead to an observable signature that would help us to distinguish between a regular neutron star and one with a color superconducting core?
These are just some of many interesting problems that need to be investigated in the future and which could have important physical consequences.

\begin{theacknowledgments}
The author thanks the organizing committee for their invitation and kind hospitality. This work was supported in part by the Office of Nuclear Physics of the Department of Energy under contract DE-FG02-09ER41599.
\end{theacknowledgments}


\bibliographystyle{aipproc}



\end{document}